\documentclass[aps,pra,reprint,showpacs,amssymb,amsmath,longbibliography]{revtex4-1}

\usepackage{graphicx}
\usepackage{times}
\usepackage{color}

\newcommand{\bn}{\begin{eqnarray}}
\newcommand{\en}{\end{eqnarray}}
\newcommand{\eml}{\end{multline}}
\newcommand{\bml}{\begin{multline}}

\begin{document}
\title{Time limited optimal dynamics beyond the Quantum Speed Limit}

\author{Miroslav Gajdacz$^{1,2}$, Kunal K. Das$^3$,  Jan Arlt$^1$, Jacob F. Sherson$^1$ and Tom\'{a}\v{s} Opatrn\'{y}$^2$}
\address{
$^1$Department of Physics and Astronomy, Aarhus University, Ny Munkegade 120, 8000 Aarhus C, Denmark\\
$^2$Optics Department, Faculty of Science, Palack\'{y} University, 17.  Listopadu 12, 77146 Olomouc, Czech Republic\\
$^3$Department of Physical Sciences, Kutztown University of Pennsylvania, Kutztown, Pennsylvania 19530, USA}

\date{\today}

\begin{abstract}
The quantum speed limit sets the minimum time required to transfer a quantum system completely into a given target state.
At shorter times the higher operation speed has to be paid with a loss of fidelity. Here we quantify the trade-off between the fidelity and the duration in a system driven by a time-varying control. The problem is addressed in the framework of Hilbert space geometry offering an intuitive 
interpretation of optimal control algorithms. This approach is applied to non-uniform time variations which leads to a necessary criterion for control optimality applicable as a measure of algorithm convergence. The time fidelity trade-off expressed in terms of the direct Hilbert velocity provides a robust prediction
of the quantum speed limit and allows to adapt the control optimization such that it yields a predefined fidelity. 
The results are verified numerically in a multilevel system with a constrained Hamiltonian, and
a classification scheme for the control sequences is proposed based on their optimizability.
\end{abstract}
\pagestyle{plain}
\maketitle

\section{Introduction}

One of the most fundamental elements of quantum mechanics is the uncertainty principle limiting simultaneous knowledge of non-commuting variables. In particular its time-energy analog investigated by Mandelstam and Tamm~\cite{JPhysMos.9.249} provides a general limit for the evolution of observables, which led Bhattacharyya~\cite{JPhysA.16.2993} to the formulation of the Quantum Speed Limit (QSL).
This principle asserts, that a system evolving from $|\psi_i\rangle$ to $|\psi_f\rangle$ in time $T$ fulfills
\mbox{$\Delta E \times T \ge \arccos(|\langle \psi_i| \psi_f \rangle|)$ }, where $\Delta E$ is the energy uncertainty of the system.

Aharonov and Anandan~\cite{PhysRevLett.65.1697} later identified $\int_0^T \Delta E dt$ with the path length of the trajectory in Hilbert space, and showed that its value is limited by $\arccos(|\langle \psi_i| \psi_f \rangle|)$. This geometrical interpretation of the QSL motivated 
Carlini~\textit{et al.}~\cite{PhysRevLett.96.060503} to search for the optimal path in Hilbert space, 
and the QSL was furthermore applied to a wide range of systems~\cite{PhysRevLett.72.3439, PhysRevA.82.022107, PhysRevLett.111.010402, PhysRevLett.110.050403, PhysRevLett.110.050402, PhysRevLett.103.160502}. 
Recently Caneva~\textit{et al.}~\cite{PhysRevLett.103.240501} demonstrated the existence of the QSL based on the convergence of an Optimum Control (OC) algorithm.

The quantum speed limit is often stated in terms of the minimum time $T = T_{QSL}$ required to obtain complete transfer into a given target state.
At durations shorter than $T_{QSL}$, the target state cannot be reached fully and the high operation speed has to be paid with a certain infidelity.
The standard QSL provides only a lower bound for $T_{QSL}$, which can be reached by an ideal Hamiltonian driving the 
system along a geodesic in Hilbert space. In most systems, however, such a Hamiltonian is not available and the actual $T_{QSL}$
is substantially larger than that lower bound. The time fidelity trade-off---a particular case of Pareto optimization~\cite{PhysRevA.78.033414}---has previously been evaluated for specific quantum systems using mainly numerical means \cite{PhysRevA.82.042305, PhysRevA.84.012312, PhysRevA.86.062309}. 
The derivative of fidelity with respect to process duration was also obtained analytically for a uniform extension of the process 
\cite{JChemPhys.130.034108, PhysRevA.85.033406}.
However, an intuitive interpretation of the trade-off as well as a treatment of non-uniform time variations has been missing.

In this article we investigate the optimality of time limited dynamics within the framework of Hilbert space geometry, where the time evolution is represented as a trajectory and the optimized quantity is the final distance from some target state.
After introduction of the basic geometrical concepts in Hilbert space, we derive a simple optimizing procedure equivalent to the standard OC algorithms.
We then examine the effect of generally non-uniform time variations, yielding a quantitative measure of process optimality, which allows to asses convergence of OC algorithms.

We express the exact time fidelity trade-off in an integral form and argue for its broad applicability in the estimation of $T_{QSL}$.
This result can also be employed in reaching a desired fidelity in a minimal time below $T_{QSL}$.
Finally, we show the existence of multiple locally optimal solutions in a system with a constrained Hamiltonian, and verify  the validity of the analytical results numerically.

\section{Hilbert space geometry}
\label{HSG}

Consider a system characterized by a state vector $|\psi\rangle \equiv |\psi(t)\rangle$ evolving in time via the Schr\"{o}dinger equation \mbox{$| \dot{\psi} \rangle = - i \hat{H}|\psi \rangle $}, where $\hat{H}$ is the time dependent Hamiltonian of the system and $\hbar = 1$.

The time derivative of the state can be interpreted as the \textit{velocity in the Hilbert space}.
Generally the \textit{parallel Hilbert velocity}
 \mbox{$|\dot{\psi}_\parallel \rangle \equiv |\psi \rangle \langle \psi | \dot{\psi} \rangle
  = -i |\psi \rangle \langle \psi|\hat{H} | \psi\rangle \equiv -i E |\psi \rangle$}
merely evolves the phase of the current state, while the \textit{perpendicular Hilbert velocity}
\mbox{$|\dot{\psi}_\perp \rangle \equiv |\dot{\psi}\rangle - |\dot{\psi}_\parallel \rangle $},
 $|\dot{\psi}_\perp| = \sqrt{|\dot{\psi}|^2 - |\dot{\psi}_\parallel|^2} = \sqrt{\langle\hat{H}^2\rangle - \langle\hat{H}\rangle^2} \equiv \Delta E$,
induces motion in the Hilbert space.

This can be seen explicitly by decomposing the state in a fixed orthonormal basis $|\phi_j \rangle$,
$|\psi\rangle = \sum_j a_j e^{-ib_j} | \phi_j \rangle$,
where $\mathbf{a} \equiv (a_1, a_2,...)$~and~$\mathbf{b} \equiv (b_1, b_2,...)$ are real vectors, and $|\mathbf{a}|^2 = \sum_j a_j^2 = 1$.
The Hilbert velocity is
\begin{equation}
  \label{vel_decomp}
  |\dot{\psi}\rangle = \sum_j  (\dot{a}_j - i a_j \dot{b}_j )e^{-ib_j}|\phi_j\rangle.
\end{equation}
At a given instant, the particular choice of basis \mbox{$|\phi_1 \rangle = |\psi \rangle $} ensures
$a_k = 0$ for $k > 1$ and $\dot{a}_1 = 0$ (since \mbox{$\frac{d}{dt} |\mathbf{a}| =0$}).
Thus a non-zero perpendicular Hilbert velocity component \mbox{$\langle \phi_{k} | \dot{\psi} \rangle = \dot{a}_k e^{-ib_k}$}
implies a time variation of the coefficient $a_k$ leading to motion in Hilbert space.

In a general basis, one finds that \mbox{$|\dot{\psi}|^2  = |\dot{\mathbf{a}}|^2 + \langle \dot{b}^2 \rangle$}
and \mbox{$| \dot{\psi}_\parallel | = \langle \dot{b} \rangle$}, where the notation $\langle c \rangle \equiv \sum_j a_j^2 c_j $ was used.
The speed of motion can then be expressed as \mbox{ $|\dot{\psi}_\perp| =  \sqrt{|\dot{\mathbf{a}}|^2 + (\Delta \dot{b})^2 }$},
where $\Delta \dot{b}\equiv \sqrt{\langle \dot{b}^2 \rangle - \langle \dot{b} \rangle^2}$.
The \textit{trajectory length} can be defined for any $|\psi(t) \rangle$, \mbox{$t\in \langle 0,T\rangle$} as
\begin{equation}
  \label{traj_len}
  \mathcal{C} \equiv \int_0^T |\dot{\psi}_\perp| dt = \int_0^T \Delta E(t) dt,
\end{equation}
which is the Aharonov-Anandan geometrical distance~\cite{PhysRevLett.65.1697}.

\begin{figure}[h!]
		\includegraphics[width=\linewidth]{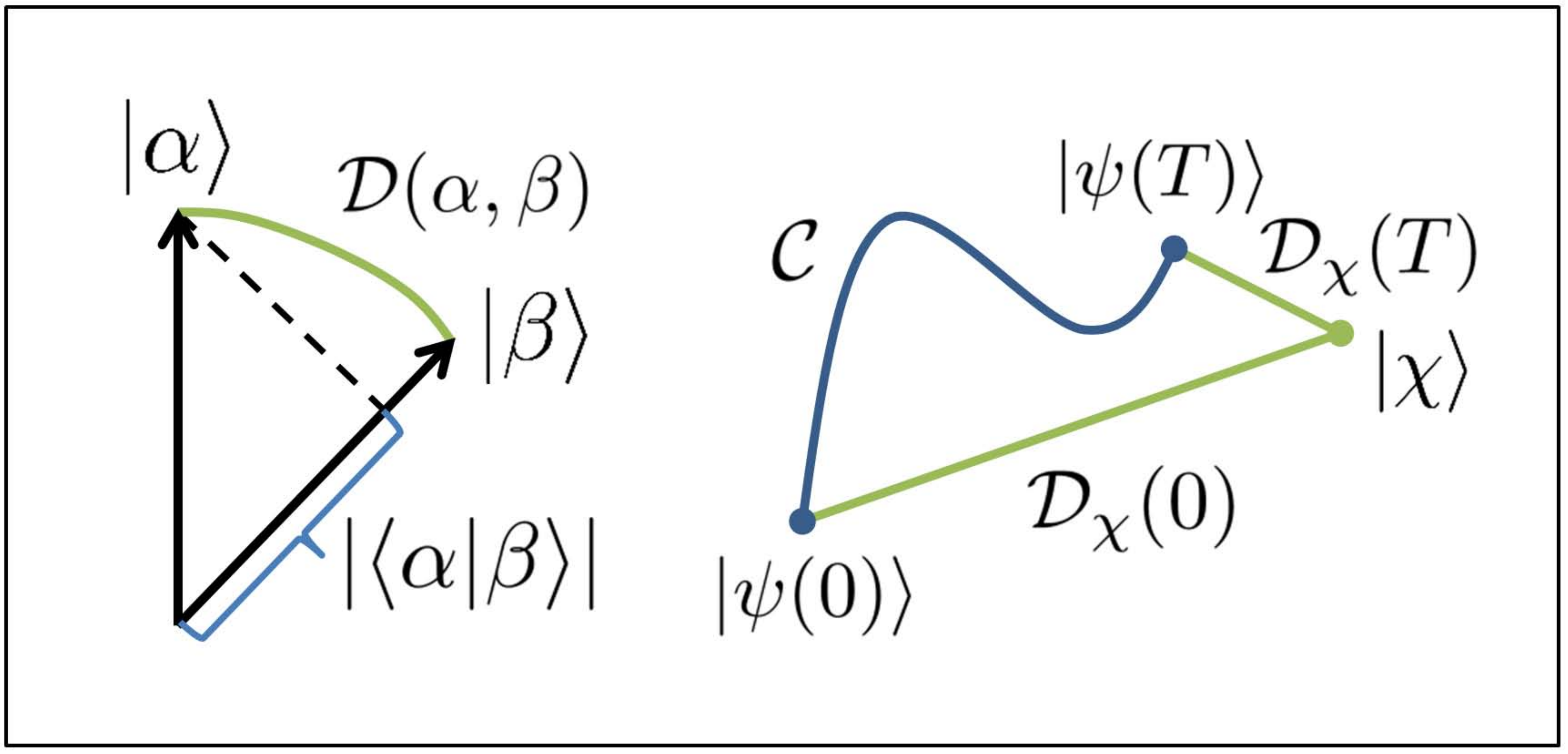}
    \caption{Schematic illustration of the distance of states from equation~(\ref{Dvalue}) and the distance inequality~(\ref{approachSpeedLimit}).}
    \label{fig:distanceIllustration}
\end{figure}
The distance in Hilbert space $\mathcal{D}(\alpha,\beta)$ between states $|\alpha \rangle$ and $|\beta \rangle$ is the length of the 
shortest trajectory connecting them. 
The above functional attains an extremal value when its integrand fulfills the Euler-Lagrange equations. 
Since $|\dot{\psi}_\perp|(\dot{\mathbf{a}},\mathbf{a},\dot{\mathbf{b}})$
does not depend on $\mathbf{b}$, the generalized momenta
\begin{equation}
  \Pi_j \equiv \frac{\partial |\dot{\psi}_\perp|}{\partial \dot{b}_j} =
  \frac{ \left(\dot{b}_j -  \langle \dot{b} \rangle \right){a_j}^2}{|\dot{\psi}_\perp|} = \mathrm{const}.
\end{equation}
are conserved.
Without loss of generality, we can choose $|\phi_1\rangle = |\alpha\rangle$ in the state expansion implying $a_1(t = 0) = 1$ and $\Pi_j = 0$ for all $j$.
At any later time, non-zero $a_j$ requires
\mbox{$\dot{b}_j = \langle \dot{b} \rangle$} and consequently $\Delta \dot{b} = 0$. In this case $|\dot{\psi}_\perp| = |\dot{\mathbf{a}}|$ for all times, and the shortest trajectory is a geodesic on a hypersphere in the space of parameter $\mathbf{a}$  defined by $|\mathbf{a}| = 1$. Identifying $|\beta\rangle$ with $|\psi (T)\rangle$, in
 the chosen basis $a_1(T) = |\langle \phi_1 | \psi(T) \rangle|=|\langle \alpha | \beta \rangle|$. Thus the distance of states is
\begin{equation}
  \label{Dvalue}
  \mathcal{D}(\alpha,\beta) = \arccos \left( |\langle \alpha | \beta \rangle| \right),
\end{equation}
which is equivalent to the Wootters distance~\cite{PhysRevD.23.357, PhysRevA.82.022107, PhysRevLett.72.3439}, and attains a maximum value $\pi/2$ for a pair of orthogonal states, see Fig.~\ref{fig:distanceIllustration}. Since $\mathcal{C} \ge \mathcal{D}$, we arrive at the integral form of the QSL inequality
\begin{equation}
 \label{bhat_gen}
 \int_0^T \Delta E dt \ge \arccos \left( |\langle \psi(T) | \psi(0) \rangle| \right).
\end{equation}
For a constant $\Delta E$ we recover the Bhattacharyya bound \mbox{$\Delta E \times T \ge \arccos \left( |\langle \psi(T) | \psi(0) \rangle| \right) $}.

\section{Relative motion}

In general, optimum control algorithms aim to drive the system into a certain 
predefined state by dynamically varying its Hamiltonian.
It is thus of special interest to evaluate the relative motion in the subspace spanned by the 
current state $|\psi\rangle$ and some fixed target state $|\chi\rangle$.
Let $|\nu\rangle$ be another fixed state forming an orthonormal basis with $|\chi\rangle$ in this subspace at a given instant.
The current state can then be expressed as
\begin{equation}
  |\psi \rangle = a_1 e^{-ib_1} |\chi\rangle + a_2 e^{-ib_2} |\nu\rangle,
\end{equation}
and the motion in the subspace is induced by a component of the perpendicular Hilbert velocity along a state
\begin{equation}
  \label{xiDef}
  |\xi \rangle = a_2 e^{-ib_1} |\chi\rangle - a_1 e^{-ib_2} |\nu\rangle
\end{equation}
(orthogonal to $|\psi\rangle$, determined up to a phase).
Defining the fidelity $F \equiv  \left| \langle \chi | \psi\rangle \right|^2 = \cos^2 \left[\mathcal{D}(\chi, \psi) \right]$, we can obtain this state from  
\mbox{$|\xi \rangle = \frac{|\chi \rangle \langle \chi | - F }{\sqrt{F(1-F)}} |\psi\rangle$}.
The states $|\psi\rangle$ and $|\xi\rangle$ also form an orthonormal basis in the subspace, hence 
\begin{equation}
  \label{chiDecompInSub}
  |\chi \rangle = e^{ib_1} \left (a_1  |\psi\rangle + a_2 |\xi\rangle \right ),
\end{equation}
implying that $|\xi\rangle$ represents the part of $|\chi \rangle$ which is not present in $|\psi\rangle$.

Using the expansion for Hilbert velocity from Eq.~(\ref{vel_decomp}) with $|\phi_1\rangle = |\chi\rangle$ and $|\phi_2\rangle = |\nu\rangle$,
we can express the perpendicular Hilbert velocity in the subspace as
\begin{align}
    |\dot{\psi}_{\perp,\chi} \rangle &\equiv |\xi \rangle \langle \xi |\dot{\psi}\rangle 
= |\xi\rangle \left[ \frac{\dot{a}_1}{a_2} + i (\dot{b}_2 - \dot{b}_1)a_1 a_2 \right],
\end{align}
where we have also used Eq.~(\ref{xiDef}) and the normalization condition \mbox{${a_1}^2 + {a_2}^2 = 1$}.
Denoting the immediate distance from $|\chi\rangle$ as $\mathcal{D}_\chi (t)\equiv \mathcal{D}(\chi,\psi(t)) = \arccos(a_1)$, 
we see that the real part of $\langle\xi|\dot{\psi}\rangle$ corresponds to the
direct motion towards the state $|\chi\rangle$
\begin{equation}
  \dot{\mathcal{D}}_\chi(t) = \frac{d}{dt} \arccos(a_1) =  - \frac{\dot{a}_1}{a_2}
= - \mathrm{Re} \langle\xi|\dot{\psi}\rangle. 
\label{directSpeed}
\end{equation}
On a Bloch sphere with $|\chi\rangle$ and $|\nu\rangle$ on the poles this corresponds to a motion along a meridian.
Similarly, the imaginary part
\begin{align}
    | \mathrm{Im} \langle\xi|\dot{\psi} \rangle | = |(\dot{b}_2 - \dot{b}_1)a_1 a_2| = 
    \sqrt{\langle \dot{b}^2 \rangle - \langle \dot{b} \rangle^2} \equiv \Delta \dot{b}
\end{align}
represents a motion along the parallels on the sphere preserving the distance from the poles.

When $|\chi\rangle$ is orthogonal to $|\psi\rangle$ we have $a_1 = 0$ and $|\xi \rangle = e^{-ib_1} |\chi\rangle$,
with an arbitrary phase $b_1$. 
The imaginary part of $\langle \xi |\dot{\psi}\rangle$ becomes zero due to vanishing frequency uncertainty $\Delta \dot{b} = 0$
and the direct velocity towards $|\chi\rangle$ becomes \mbox{$\dot{\mathcal{D}}_\chi(t) = - |\langle\chi|\dot{\psi}\rangle|$}.

For a general trajectory $|\psi(t) \rangle$, $t \in \langle 0, T \rangle$ we can obtain the distance of its end point from the target by
integrating Eq.~(\ref{directSpeed}) 
\begin{equation}
   \label{distanceIntegral}
    \mathcal{D}_\chi(T) = \mathcal{D}_\chi(0) - \int_0^T \mathrm{Re}  \langle\xi(t)|\dot{\psi}\rangle dt,
\end{equation}
where the time dependence of $|\xi\rangle$ was shown explicitly.
Since $\langle\xi|\dot{\psi}\rangle$ is only one component of the transverse Hilbert velocity, it directly follows 
\begin{equation}
   \label{approachSpeedLimit}
   \mathcal{D}_\chi(T) \ge \mathcal{D}_\chi(0) - \int_0^T \Delta E dt,
\end{equation}
which can also be seen by realizing that the hypothetical trajectory $\mathcal{C} + \mathcal{D}_\chi(T)$ connecting  $|\psi(0)\rangle$
and $|\chi\rangle$ is necessarily longer or equal to the distance of the two states $\mathcal{D}_\chi(0)$, see Fig.~\ref{fig:distanceIllustration}. 
The above expression sets a limit on how quickly a target state can be approached 
as opposed to Eq.~(\ref{bhat_gen}), which sets a limit on how quickly a system can leave an initial state.

\section{Optimal navigation}

Just like it often pays off to take a slightly longer path to avoid an obstacle on the way to our goal,
it may not be optimal to maximize the direct Hilbert velocity towards the target at all times. 
Taking a longer path at higher speed may produce a better result.
What is important is the final proximity to the target achieved in the specified time, rather than the actual traveled distance.

In the following we will consider a case when the Hamiltonian of the system depends on time via a vector of 
control parameters ${\bf u}(t)$, that is $\hat{H}\equiv \hat{H}({\bf u}(t))$. Suppose the initial state $|\psi(0)\rangle$ is fixed and
 we have some guess for the control ${\bf u}(t)$, $t \in \langle 0, T \rangle$. To obtain the final distance from the target  
$\mathcal{D}_\chi(T)$, we first need to calculate the full time evolution of the initial state. 
How will  $\mathcal{D}_\chi(T)$ change when we arbitrarily alter the control on some short time interval within the process?

Thanks to unitarity of the quantum time evolution we do not have to  calculate the whole trajectory again:
For any two trajectories $|\psi_1(t)\rangle$ and $|\psi_2(t)\rangle $ governed by 
the same Hamiltonian and having generally different starting points
$|\psi_1(0)\rangle$ and $|\psi_2(0)\rangle$, the immediate distance $\mathcal{D}\left(  \psi_1(t), \psi_2(t) \right)$ is 
preserved for all times $t$. 
This follows from the time invariance of the scalar product
\begin{align}
\nonumber
\frac{d}{dt} \langle \psi_1  |\psi_2 \rangle &= \langle \dot{\psi}_1  |\psi_2 \rangle + \langle \psi_1  |\dot{\psi}_2 \rangle \\
   &= i \langle \psi_1| \hat{H}  |\psi_2 \rangle - i \langle \psi_1 |\hat{H}  |\psi_2 \rangle = 0.
\end{align}

\begin{figure}[!h]
		\includegraphics[width=\linewidth]{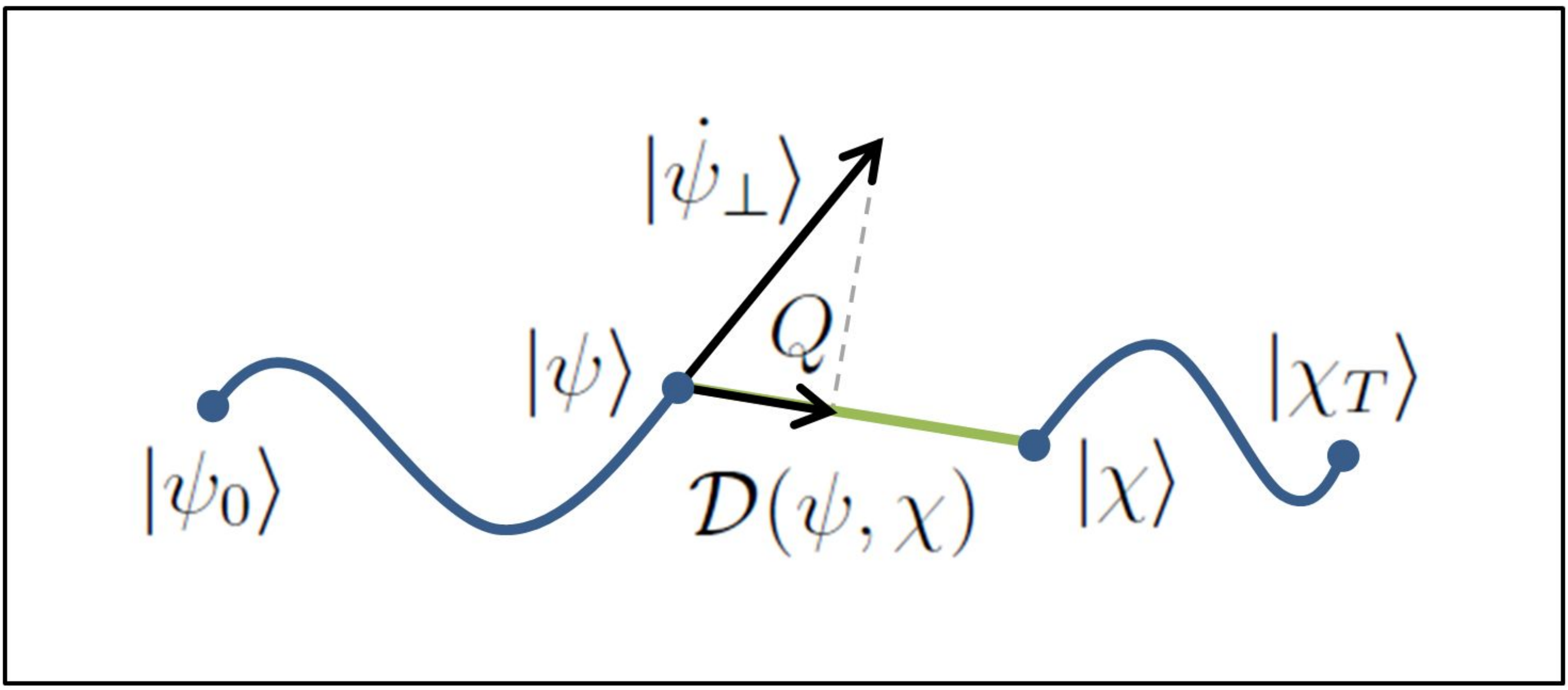}
    \caption{Schematic interpretation of the direct Hilbert velocity $Q$ as a component of the perpendicular Hilbert velocity $|\dot{\psi}_\perp\rangle$ which 
corresponds to the shortening rate of the distance $\mathcal{D}(\psi,\chi )$, see Eq.~(\ref{DistanceDep}). We have omitted the explicit time dependence of the forward evolved initial state $|\psi \rangle \equiv |\psi(t) \rangle$, as well as the backwards evolved target state $|\chi \rangle \equiv |\chi(t)\rangle$. 
Subscripts on $\psi$ and $\chi$ denote points in time.}
    \label{fig:Qinterpretation}
\end{figure}

For convenience of notation we will rename the target state $\chi \rightarrow \chi(T)$ and denote its 
backwards evolved trajectory as $|\chi(t)\rangle$. The final distance from the target is then equal to the immediate distance
of the trajectories $|\psi(t)\rangle$ and $|\chi(t)\rangle$
\begin{equation}
   \mathcal{D}_\chi(T) \equiv \mathcal{D}\left(  \chi(T), \psi(T) \right) = \mathcal{D}\left(  \chi(t), \psi(t) \right)
\end{equation}
for any point in time. Utilizing the result~(\ref{distanceIntegral}) for infinitesimal integration boundaries $\langle t - dt, t\rangle$, 
we can write 
\begin{equation}
    \label{DistanceDep}
    \mathcal{D}_\chi(T) =  \mathcal{D}\left(  \chi(t), \psi(t - dt) \right) - Q(t) dt,
\end{equation} 
where we have introduced a new notation for the \emph{direct Hilbert velocity} 
\begin{equation} 
    Q(t) \equiv \mathrm{Re} \langle\xi(t)|\dot{\psi}\rangle = \mathrm{Im} \langle\xi(t)|\hat{H}(t)|\psi(t)\rangle.
\end{equation} 
Note that the state $|\xi(t)\rangle$ is now computed with respect to the backwards evolved target state $|\chi(t)\rangle$.
As before, the direct Hilbert velocity is bounded from above by
\begin{equation}
  \label{QasCompDeltaE}
  Q \equiv \mathrm{Re} \langle \xi |\dot{\psi} \rangle = \mathrm{Re} \langle \xi |\dot{\psi}_\perp \rangle \le |\dot{\psi}_\perp| = \Delta E.
\end{equation}
The equality occurs when the motion in the Hilbert space is along a geodesic towards $|\chi(t)\rangle$.

Equation~(\ref{DistanceDep}) shows that in order to minimize the final distance from the target, we have to maximize the direct
Hilbert velocity $Q(t)$ at each point in time. The simplest local optimization algorithm can vary the control proportionally to the gradient of $Q(t)$
\begin{equation}
  \delta {\bf u}(t) = \alpha \cdot \frac{\partial Q(t)}{\partial {\bf u}} = \alpha \cdot \mathrm{Im} \left \langle\xi(t) \left |\frac{\partial \hat{H}}{\partial {\bf u}} \right|\psi(t) \right\rangle,
\end{equation}
with some step size $\alpha$. 
An improved convergence can be achieved by employing higher derivatives with respect to the control \cite{PhysRevA.83.053426}.
Once the control has been altered on some finite time interval, one can update the time evolution of 
$|\psi(t)\rangle$ and $|\chi(t)\rangle$ on that interval and proceed by optimizing a neighboring interval
 (preceding or following in time) \cite{NewJPhys_13_073029}. 
One iteration of the algorithm would then be understood as a sweep over the whole process duration.

Such an optimization is in fact equivalent to the Krotov algorithm~\cite{PhysRevA.66.053619, GlobMet_Krotov}
which follows from the Pontryagin maximum principle~\cite{PhysRevA.65.032301, Pontryagin} 
as well as alternative approaches~\cite{Khaneja2005296, PhysRevA.72.023416}.
In the Krotov algorithm the optimized quantity is fidelity and the improvement of the control
is found by maximization of the Pontryagin Hamiltonian $\mathcal{H}(t,\psi,{\bf u},\chi) 
\equiv 2\mathrm{Im} \left[ \langle \chi | \hat{H} |\psi\rangle \langle \psi|\chi \rangle \right]$.
Inserting for $|\chi\rangle$ from Eq.~(\ref{chiDecompInSub}), we see that this function is in fact
proportional to the direct Hilbert velocity
\begin{equation}
   \mathcal{H}(t,\psi,{\bf u},\chi) = 2\sqrt{F(1-F)} \times Q(t).
\end{equation}
Thus our result provides an interpretation of optimum control theory in terms of
Hilbert space geometry. As shown below it also offers an intuitive framework for the understanding of time optimization.

\section{Time fidelity trade-off}

We now turn to the question of trade-off between the duration of the process and the achievable proximity of the target state. 
To our knowledge, this problem has only been studied for uniform extensions of the 
process~\cite{PhysRevA.78.033414,JChemPhys.130.034108, PhysRevA.82.042305, PhysRevA.84.012312, PhysRevA.85.033406, PhysRevA.86.062309}.
In the following we consider a more general case of non-uniform time variations.

Assume the process can be divided into $N$ small but finite time intervals $\Delta t_j$, connected at points in time \mbox{$t_j = \sum_{k = 1} ^j \Delta t_k$}.
At each interval the Hamiltonian is constant and determined uniquely by the value of the control parameter 
${\bf u}_j $, thus the set $\{{\bf u}_j, \Delta t_j; j=1,2,...,N\}$, together with the initial condition for $|\psi(0)\rangle$,
completely defines the process. 
 
When treating $\Delta t_j$ as independent parameters, the process duration $T$ is also allowed to vary, however both 
$|\psi(0)\rangle$ and $|\chi(T)\rangle$ remain fixed.
A general variation of the time intervals can be written in the form $\delta \Delta t_j = \mu_j \Delta t_j$, where all $|\mu_j| \ll  1$. 
To the first order in $\Delta t_j$ we can approximate Eq.~(\ref{DistanceDep}) as 
\begin{equation}
    \label{DistanceDepDiscrete}
    \mathcal{D}_\chi(T) \approx  \mathcal{D}\left( \chi(t_j), \psi(t_{j-1}) \right) - Q_j \Delta t_j,
\end{equation} 
with $Q_j \equiv Q(t_j)$.
The induced variations of $T$ and $\mathcal{D}_\chi(T)$ then are
\begin{align}
  \delta T = \sum_{j=1}^N \delta \Delta t_j = \sum_{j=1}^N \mu_j \Delta t_j &= T \left\langle \mu \right\rangle_T, \\
\label{fid_variation}
  \delta \mathcal{D}_\chi(T) = \sum_{j=1}^N \frac{\partial \mathcal{D}_\chi(T)}{\partial \Delta t_j} \delta \Delta t_j &=
 - T \left\langle Q \mu \right\rangle_T,
\end{align}
where we have defined the time average 
\begin{equation}
   \langle f \rangle_T \equiv \frac{1}{T} \sum_{j=1}^N f_j \Delta t_j  \rightarrow 
\frac{1}{T} \int_0^T f(t) dt.
\end{equation}

For the case of an  \textit{uncorrelated adjustment} $\mu_j$, fulfilling 
\begin{equation}
   \label{CovDef}
    \mathrm{Cov}\left(Q, \mu \right) \equiv \left\langle Q \mu \right\rangle_T -  
     \left\langle Q \right\rangle_T    \left\langle \mu \right\rangle_T = 0,
\end{equation}
the variation of the distance is simply
\begin{equation}
\label{dDdT}
  \delta \mathcal{D}_\chi(T) = - \left\langle Q\right \rangle_T \delta T.
\end{equation}
A trivial example fulfilling condition~(\ref{CovDef}) is a uniform extension of the process \mbox{$\delta \Delta t_j = \kappa \Delta t_j$}, with a small constant 
\mbox{$\kappa = \delta T /T$}. This case was considered among others by Mishima~\textit{et al.}~\cite{JChemPhys.130.034108} 
arriving at an equivalent time fidelity trade-off, which in our notation can be expressed as
\begin{align}
\label{dFdT}
  \frac{dF}{dT} &= \frac{1}{T} \int_0^T 2\mathrm{Im} \left[ \langle \chi | \hat{H} |\psi\rangle \langle \psi|\chi \rangle \right] dt \\
                      &= 2\sqrt{F(1-F)}\times \langle Q \rangle_T,
\end{align}
where the $|\chi \rangle$ decomposition~(\ref{chiDecompInSub}) was utilized.

Let us now consider a generally non-uniform adjustment of the time intervals which preserves the total duration $T$. 
Such a redistribution must be of the form  
\mbox{ $\mu_j = \epsilon \left[ \nu_j - \left\langle \nu \right\rangle_T \right]$}, where $\nu_j \equiv \nu(t_j)$ is an arbitrary function of time, and
$\epsilon$ is a small scaling factor.
Using Eq.~(\ref{fid_variation}), the corresponding change in the distance is
\begin{equation}
\delta \mathcal{D}^\epsilon_\chi(T) =  -\epsilon T\ \mathrm{Cov}\left(Q, \nu  \right),
\end{equation}
which is extremal for $\nu_j = Q_j$. Comparing this with
 the distance variation $\delta \mathcal{D}^\kappa_\chi(T)$ induced by a uniform extension of the process with an equivalent mean adjustment 
\begin{equation}
   \kappa = \sqrt{\left\langle {\mu}^2 \right\rangle_T} = \epsilon \sqrt{\langle (Q - \langle Q \rangle_T )^2 \rangle_T } \equiv \epsilon\ \mathrm{Std}\left(Q\right),
\end{equation}
we obtain a measure of the process optimality 
\begin{equation}
\label{conv_crit}
\sigma_Q \equiv \frac{\delta \mathcal{D}^\epsilon_\chi(T)} {\delta \mathcal{D}^\kappa_\chi(T)} = 
 \frac{ \mathrm{Cov}\left(Q,Q \right)}{ \mathrm{Std}\left(Q\right) \langle Q \rangle_T } = 
 \frac{\mathrm{Std}\left(Q\right) }{ \langle Q \rangle_T } .
\end{equation}

For a sufficiently fine discretization of time, any process optimal with respect to $\mathbf{u}_j$ is necessarily 
extremal with respect to any variation of $\Delta t_j$ which preserves $T$, implying $\delta \mathcal{D}^\epsilon_\chi(T) \rightarrow 0$.
Thus $\sigma_Q \rightarrow 0 $ is a necessary criterion for process optimality, and can be used for quantifying the convergence of OC algorithms.
Additionally $\sigma_Q = 0$ implies  
$Q(t) = \langle Q \rangle_T $ for all points in time, which via 
\mbox{$\mathrm{Cov}\left(Q, \mu  \right) = 0$} guarantees validity of Eq.~(\ref{dDdT}) 
for any time adjustment $\mu_j$ of an optimal process. 

For further discussion it is useful to introduce a classification scheme of the control sequences based on their optimality. Since the optimizing algorithm searches for local optima, the optimization result can depend on the initial choice of the control ${\bf u}(t)$. We define an \textit{optimum class} as a continuous $T$ transformation of \textit{optimal} control parameters $\mathbf{u}_{\mathrm{opt}}(T, t) \equiv \mathbf{u}_{\mathrm{opt},T}(t)$. 
A set of initial control parameters yielding upon optimization a solution in a certain optimum class will be called a \textit{control family}.

If we denote the direct Hilbert velocity within an optimum class by $Q_\mathrm{opt}(T)$, we can write the time distance trade-off~(\ref{dDdT})
in an integral form
\begin{equation}
  \label{dist_trade}
  \mathcal{D}_\chi(T_2) = \mathcal{D}_\chi(T_1) - \int_{T_1}^{T_2} Q_\mathrm{opt}(T) dT.
\end{equation} 
For an optimum class extending from zero to some finite duration $T$, the above equation quantifies the speed limit exactly
as opposed to Eq.~(\ref{approachSpeedLimit}), which merely provides a lower bound.
In terms of fidelity the above can be written as
\begin{equation}
  \label{fid_trade}
  \left[ \arcsin \left( \sqrt{F} \right) \right]^{F_2}_{F_1} = \int_{T_1}^{T_2} Q_\mathrm{opt}(T)  dT.
\end{equation}

Usually the convergence of OC algorithms becomes slower as $T$ approaches the quantum speed limit $T_\mathrm{QSL}$
from below. Interestingly for many systems $Q_\mathrm{opt}(T)$ is constant or a slowly varying function of $T$ in that regime.
The value of $T_\mathrm{QSL}$ can thus be predicted well 
 even for moderate values of fidelity ($F \approx 0.9$, $T < T_\mathrm{QSL}$) by approximating the integrand in Eq.~(\ref{fid_trade}) with a constant.
Note that Eq.~(\ref{dFdT}) is not very suitable for linear extrapolation of the fidelity, since the right hand side varies quickly when $F \rightarrow 1$
and thus cannot be approximated with a constant. 

Caneva~\textit{et al.}~\cite{PhysRevA.84.012312} observed the relation $F = \sin^2\left(\frac{\pi}{2} T/T_{QSL} \right)$ 
arising from a numerical optimization of multiple physical systems, and attributed this behavior to the motion along geodesics in Hilbert space. 
In general, Eq.~(\ref{fid_trade}) implies 
$F = \sin^2\left ( \int_0^T Q_\mathrm{opt} (T^\prime) dT^\prime \right)$
for an optimum class with $F(T=0) = 0$.
The $\sin^2$ dependence thus occurs whenever $ Q_\mathrm{opt}(T)$ is independent of $T$, even if the motion is not along a geodesic. 
Unit fidelity is then reached in time $T_\mathrm{QSL} = \pi/(2Q_\mathrm{opt}) $.

Equation~(\ref{fid_trade}) also allows an OC algorithm to search for a process yielding a certain predefined fidelity while having the shortest possible duration within a given control family. After the default OC algorithm has converged to some fidelity $F_1$ for a given initial duration $T_1$, we can estimate the time $T_2$ required to obtain fidelity $F_2$ by setting $Q_\mathrm{opt}(T)$ constant in Eq.~(\ref{fid_trade}),
that is
\begin{equation}
  \label{T2guess}
  T_2 =  T_1 +  \left. \left[\arcsin \left( \sqrt{F} \right) \right]^{F_2}_{F_1}  \right /Q_\mathrm{opt}(T_1) .
\end{equation}
Re-optimizing the process with uniformly extended control to $T = T_2$  and repeating the estimate of $T_2$ 
converges upon the process with the desired fidelity $F_2$ in few iterations.
When $Q_\mathrm{opt}(T)$  is a varying function, we can improve the convergence by employing its derivatives in the expansion of the integrand in Eq.~(\ref{fid_trade}).\\

\section{Application to Entanglement generation in a multilevel system} 

To provide a non-trivial example of our time optimal control, we optimize entanglement generation in an atomic system with Rydberg excitation blockade~\cite{PhysRevLett.100.170504}. The system consists of $N$ indistinguishable atoms, each having two ground states $|1\rangle$ and $|2\rangle$, and a highly excited Rydberg state $|r\rangle$. The ground states are coupled by a resonant external field with a Rabi frequency $\Omega_{1}(t) = \Omega_\mathrm{max}u_1(t)$, and similarly the states $|2\rangle$ and $|r\rangle$ are coupled by $\Omega_{r}(t)  = \Omega_\mathrm{max}u_r(t)$, with 
the control parameters limited by $ 0 \le u_i \le 1$ and $\Omega_\mathrm{max} = 2\pi \times 10 \mathrm{MHz}$. Due to a large electric dipole moment, a single Rydberg excitation will render $\Omega_{r}$  off-resonant for the remaining atoms, thus permitting only one Rydberg excitation at a time. Consequently, the system is closed in the $2N + 1$ dimensional Hilbert space with a symmetric basis $|n_1,n_2,n_r\rangle$, where $n_i$ is the number of atoms in the state $|i\rangle$, and $n_1 + n_2 + n_r = N$, $n_r \le 1$. The Hamiltonian is $H = H_{J_x} + H_{\rm JC}$ with
\begin{align}
  \label{ham_Jx}
  H_{J_x}(t) &\equiv \Omega_{1}(t) J_x = \Omega_{1}(t) \frac{1}{2}\left(a_1^\dagger a_2 + a_1 a_2^\dagger \right),~~~\\
  \label{ham_JC}
  H_\mathrm{JC}(t) &\equiv \Omega_{r}(t) \frac{1}{2}\left(a_2^\dagger \sigma^- + a_2\sigma^+ \right),
\end{align}
where $a_i$ $(a_i^{\dagger})$ are the conventional annihilation (creation) operators, $J_x$ is the pseudo-spin operator and $\sigma^\pm$ are the Pauli matrices denoting the transfer between the  states with 0 and 1 Rydberg excitation.

Initially the system is prepared in $| \psi(0) \rangle = |N,0,0 \rangle$. Motivated by Ref.~\cite{PhysRevLett.100.170504}, we aim to prepare the maximally entangled state
\begin{equation}
  | \chi(T) \rangle = \left\{ \begin{array}{ll}
         |J_x = 0\rangle & \mbox{if $N$ is even}\\
         \left( |J_x = 0\rangle \otimes |r \rangle \right)_\mathrm{sym} & \mbox{if $N$ is odd}\end{array} \right. ,
\end{equation}
where $(\cdot)_\mathrm{sym}$ denotes symmetrization with respect to all atoms.
To have a simple but non-trivial system with $\langle \psi(0) | \chi(T) \rangle = 0$, we have chosen $N = 3$.

\begin{figure}[!h]
		\includegraphics[width=\linewidth]{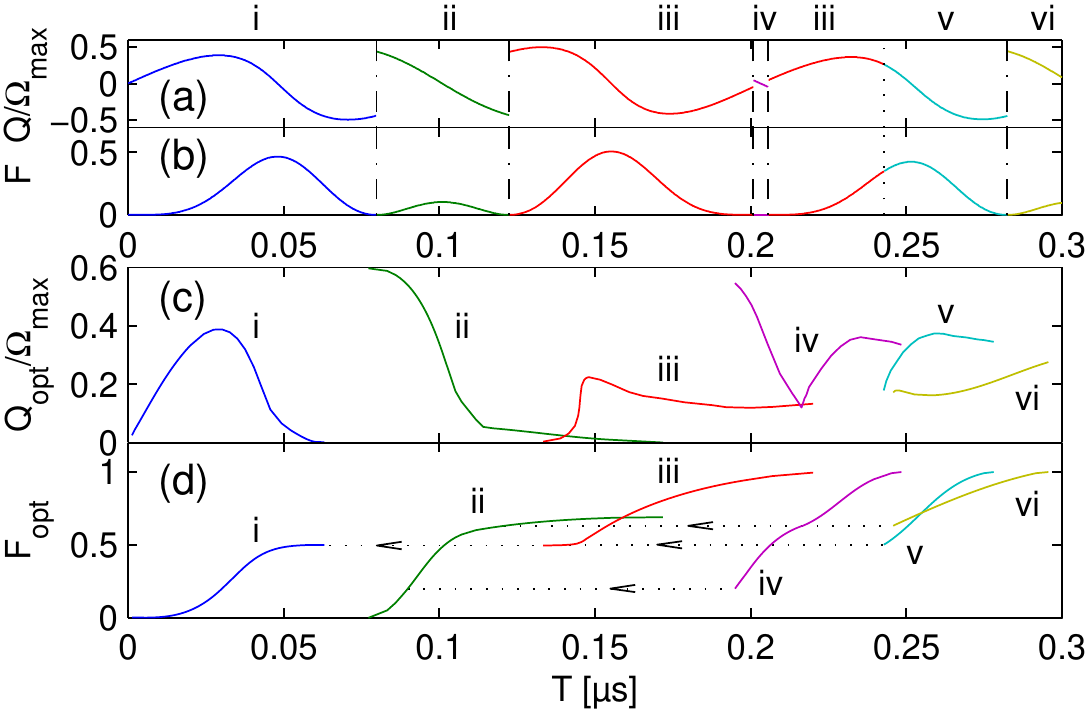}
    \caption{(a,b) The evolution of the direct Hilbert speed $Q$ and fidelity $F$ are shown for constant control \mbox{$u_1(t) = u_r(t) = 1$} as a function of the process duration $T$. Note the discontinuities in $Q(T)$ separating the control families \textit{i - vi}. (c,d) The values of $Q$ and $F$ for the optimum solutions (classes) are shown as a function of the process duration $T$. The horizontal dotted lines in sub-figure (d) represent the slip transitions (see text).}
    \label{fig:FvsT}
\end{figure}

To classify the control sequences as outlined above, we initially choose constant control parameters \mbox{$u_1(t) = u_r(t) = 1$} and
evolve the states $\psi$ and $\chi$ in time (forward and backward respectively) for a variable total duration $T$. The resulting fidelities $F(T)$ and the values of $\langle Q \rangle_T$ are shown in Fig.\ref{fig:FvsT}(a,b). Note that $Q$ does not depend on $t$ since $H$ is constant and thus commutes with the evolution operator. The examined range of $T$ is divided into several sections by discontinuities in $Q(T)$ where $F \rightarrow 0$ and $Q(T)$ changes sign.

To identify the associated control families we perform control optimization for initial parameters $u_i$ chosen from each of these sections.
Once the control has been optimized, an element in the optimum class is found and the whole class can be mapped out by allowing the process duration to vary trough Eq.~(\ref{T2guess}), where the target fidelity is adjusted in small steps. Different initial conditions converging into the same optimum class then belong to the same control family.
 The division of our initial controls into control families is denoted by roman numbering \textit{i - vi}. Note that initial control parameters from different sections can belong to the same control family as illustrated by family \textit{iii}. Moreover the transition to a different control family can occur at non-zero fidelity, illustrated by family \textit{v}.

\begin{figure}[t]                
		\includegraphics[width=\linewidth]{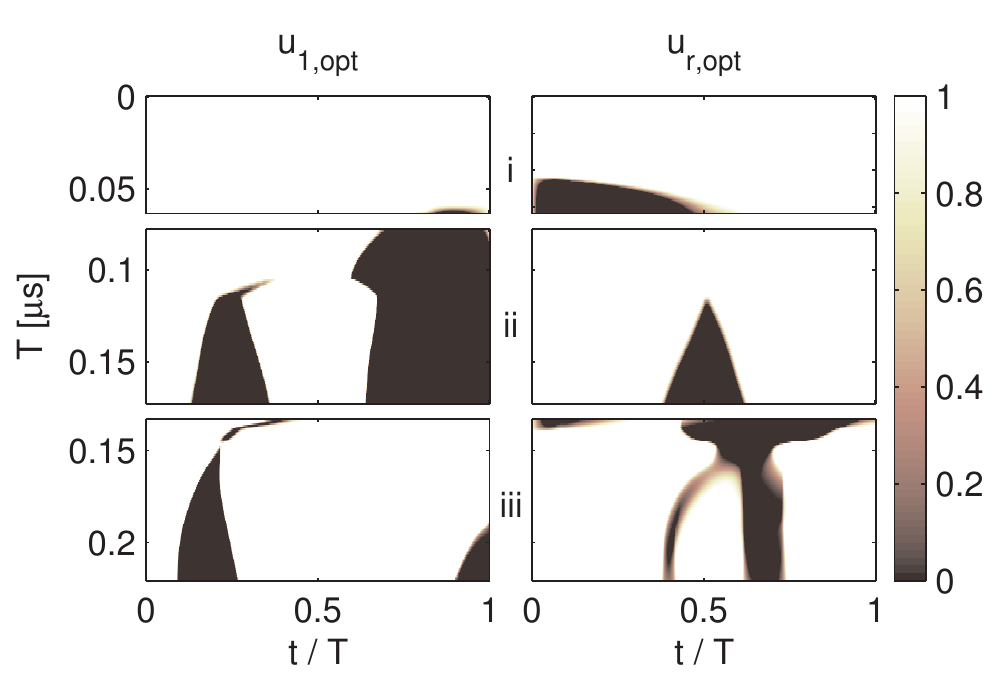}
               \caption{Time dependence of the control parameters within the optimal classes $i$, $ii$ and $iii$ referenced in Fig.~\ref{fig:FvsT}. The images show values of the control parameters $u_1$ and $u_r$ at each instant along the process $t/T$ (horizontal axis) for each process duration $T$ (vertical axis). Shading encodes the value. 
A horizontal cross section corresponds to a single optimum control sequence.}
              \label{fig:OptExamp}
\end{figure}   

Figure~\ref{fig:FvsT}(c,d) shows the fidelity $F_\mathrm{opt}(T)$ and $Q_\mathrm{opt}(T)$ for the six optimum classes. Each class is shown within the relevant region in $T$ where $0 < F < 1$ and $Q_\mathrm{opt}(T)>0$. The first two classes do not reach $F = 1$, because the OC algorithm fails to improve when $Q_\mathrm{opt}(T) \rightarrow 0$. Interestingly, the remaining optimal classes slip into a lower class before reaching $F = 0$ (denoted by vertical dotted lines in Fig.\ref{fig:FvsT}(d)). 

These slip transitions are very sudden due to the use of the modified OC algorithm aiming for some predefined fidelity. A slight decrease of the target $F$ at the slip point allows to shorten $T$ substantially by falling into a different control family and converging towards an optimal solution there. We never observe such transitions while increasing the target $F$, just as it is not possible to find the upper optimum class when extending the duration in fixed steps and optimizing the control.

Within the numerical precision of our model, all curves in Fig.~(\ref{fig:FvsT}) are consistent with Eq.~(\ref{fid_trade}).
A very important feature is the slow variation of $Q_\mathrm{opt}$ as $F \rightarrow 1$. This property 
allows us to extrapolate the fidelity in a wide range of durations and to predict the value of $T_\mathrm{QSL}$ using equation~(\ref{fid_trade}).
Thus the time fidelity trade-off can be quantified even for moderately optimized processes.

Figure~\ref{fig:OptExamp} presents the optimal control sequences for the relevant range of process durations in the optimum classes \textit{i, ii} and \textit{iii} referenced in Fig.~\ref{fig:FvsT}. Note that the function $\mathbf{u}_{\mathrm{opt}}(T,t)$ is pulse-like but continuous in both dimensions. 
This demonstrates that for small time variations the process remains close to optimal. Although some optimum classes overlap in time, they are
clearly using different strategies to approach the target.  

The presented optimum classes are not the only possible solutions to the problem, but they provide very efficient processes reaching perfect fidelity in $T_\mathrm{QSL} = 0.2204 \mathrm{\mu s}$ for the \textit{iii} class (on the order of the coupling period
$2\pi / \Omega_\mathrm{max} = 0.1 \mathrm{\mu s}$). Nevertheless, the motion in the Hilbert space is most certainly not along a geodesic, since the corresponding path length \mbox{ $\mathcal{C} = \int_0^T \Delta E dt = 10.16$} is much longer than the distance of states \mbox{ $\mathcal{D}(\psi(0),\chi(T)) = \pi/2$}.  This is due to the character of the  Hamiltonians~(\ref{ham_Jx})~and~(\ref{ham_JC}), which do not provide the ideal driving of the system.

Although this numerical example considers a finite dimensional system, the formalism is universal and applicable to any quantum system for which the state evolution can be computed.

\section{Conclusion}

In summary, we have derived a simple algorithm for local Hilbert space trajectory optimization based on Hilbert velocity analysis and demonstrated its equivalence with standard OC algorithms. Subsequently we have quantified the trade-off between the fidelity and the duration of a general process driven by a time varying control, and derived a necessary convergence criterion applicable to local OC algorithms, Eq.~(\ref{conv_crit}).  
Rather than providing a lower bound on the duration of the state evolution, as in the standard QSL criterion, equation~(\ref{dist_trade}) evaluates the speed limit exactly. In practice, equation~(\ref{fid_trade}) allows to adapt an OC algorithm to minimize the process duration while obtaining a predefined fidelity 
and to extrapolate the value of the quantum speed limit $T_\mathrm{QSL}$ from optimal processes with $T <T_\mathrm{QSL}$. The formalism developed here has broad applicability to quantum optimization problems; we illustrate this by applying it to a multilevel system with a constrained Hamiltonian, for which we present and classify a number of different optimal solutions.

\begin{acknowledgments}
This work was supported by the Lundbeck Foundation and the Danish National Research Council. T.O. and M.G. acknowledge the support of the Czech Science Foundation, grant No. GAP205/10/1657. K.K.D. acknowledges support from the National Science Foundation under Grant No. PHY-1313871
 and of a PASSHE-FPDC grant and a research grant from Kutztown University.
\end{acknowledgments}

\bibliography{OptConQSL}

\end{document}